\documentstyle[12pt,epsf]{article}

\newcommand{\be}{\begin{equation}}
\newcommand{\ee}{\end{equation}}
\newcommand{\bea}{\begin{eqnarray}}
\newcommand{\eea}{\end{eqnarray}}
\newcommand{\bean}{\begin{eqnarray*}}
\newcommand{\eean}{\end{eqnarray*}}
\newcommand{\lp}{\left(}
\newcommand{\rp}{\right)}

\newcommand{\im}{\imath}

\newcommand{\bmv}[1]{\mbox{\boldmath $#1$}}

\title{\bf Enskog-Landau kinetic equation. Calculation of the transport
	coefficients for charged hard spheres}
\author{A.E.Kobryn, V.G.Morozov$^\dag$, I.P.Omelyan, M.V.Tokarchuk
   	\\ [1.5ex]
   	\it Institute for Condensed Matter Physics\\
   	\it of the National Ukrainian Academy of Sciences\\
   	\it 1~Svientsitskii St., UA--290011 Lviv, Ukraine\\ [2ex]
	\it $^\dag$Moscow Institute of Radioengineering,\\
	\it Electronics and Automation, Physics Department\\
	\it 78 Vernadsky Av., 117454 Moscow, Russia}
\date{\today}

\begin{document}
\maketitle

\begin{abstract}
Using charged hard spheres model as an example, the dense one-com\-po\-nent 
plasma is considered. For this model the Enskog-Landau kinetic equation is
obtained and its normal solution is found using Chapman-Enskog method.
Transport coefficients are obtained numerically and analytically and
compared with the experimental data available.
\end{abstract}
\newpage

\section{Introduction}

Construction of kinetic equations for dense gases and plasma is one of the
most important problem in the kinetic theory of classical systems.
A consequent approach for construction of kinetic equations has been done by
Bogolubov \cite{c1}. This approach is based on a chain of equations for
$s$-particle distribution functions and on boundary conditions for weakening
correlations. Using such boundary conditions, we can in principle, express
all $s$-particle distribution functions in terms of the single-particle
function and obtain for it a closed kinetic equation. There is a large
number of approaches for derivation of kinetic equations \cite{c2,c3,c4,c5}.
Despite a difference of these approaches between themselves in shape, the
weakening correlation principle in one or another form has been used in all
approaches just as in Bogolubov's method. However, all these approaches are
most efficient in the case when a small parameter (density, interaction,
etc.) is present. For dense gases and dense plasma small parameters are
absent. In this case analysis of the BBGKY hierarchy becomes very difficult
because we can not restrict ourselves to some finite number of terms in
expansion for the collision integral. Moreover, an additional problem,
concerning correct account in the collision integrals of interactions
between particles on short as well as long distances, arises.

Relatively recently an approach, based on a modification of the weakening
correlation principle, has been proposed \cite{c6}. This approach leads to
a consequent construction of kinetic equations for dense gases without
additional phenomenological assumptions. New boundary condition to the
BBGKY hierarchy take into account a non-equilibriumnes of single particle
distribution function as well as local conservation laws of mass, momentum
and energy, i.e., the quantities which constitute the basic for the
hydrodynamic description of evolution of the system. In the ``pair
collision'' approximation, such approach leads to an Enskog-like kinetic
equation. Similar ideas have been proposed independently by Karkheck, van
Beijeren, de Schepper and Stell \cite{c7} at derivation of the kinetic
equation for the ``square-well'' potential. Somewhat different modification
of Bogolubov's approach has been considered by Rudyak \cite{c8,c9}. Here,
the Enskog-like kinetic equation for a system of hard spheres has been
obtained and attempts to extend this equation on a system with soft potential
have been made.

The ideas of work \cite{c6}, which is based on Zubarev's non-equilibrium
statistical operator method \cite{c10,c11}, stimulate a revision of the
problem connected with constructing of kinetic equations for dense gases
and plasma. Investigations \cite{c12,c13} were logical continuation of
the work \cite{c6} in which original result has been achieved: a consequent
derivation of the kinetic equation of revised Enskog theory
\cite{c14,c15,c16} for a system of hard spheres.

In the present paper a kinetic equation for the single-particle distribution
function is obtained from the BBGKY hierarchy with modified boundary
condition in the ``pair collision'' approximation. This kinetic equation is
valid for moderately dense classical systems with the interparticle potential
in a form
of hard sphere potential plus some long-range potential ${\mit\Phi}^l(r)$.
In the case when ${\mit\Phi}^l(r)$ is the Coulomb potential, we have
obtained a kinetic equation, called  Enskog-Landau one, for a system of
charged hard spheres. Normal solutions of this equation are found by the
Chapman-Enskog method. On the basis of the solutions, analytical expressions
for viscosity coefficients and thermal conductivity are obtained. Numerical
calculations of transport coefficients are performed for neutral and ionized
argon. The results are presented in a temperature dependent form. A
comparison between theoretically predicted values for transport coefficient
and experimental data is examined.

\section{Enskog-Landau kinetic equation}

	The BBGKY hierarchy of equations for non-equilibrium distribution
functions of classical interacting particles has been obtained in the paper
\cite{c6}
on the basis of assembling time retarded solutions for Liouville equation
with modified Bogolubov's condition meaning weakening correlations
between particles. According to Zubarev's non-equilibrium statistical operator
method \cite{c10,c11}, full non-equilibrium distribution
function $\rho\lp x^N;t\rp$ for all $N$ particles of the system satisfies
the following asymptotic condition:
%
%	1
%
\be
\lim _{t_0\rightarrow -\infty }\exp \left( \imath L_Nt_0\right) \left( \rho
\left( x^N,t_0\right) -\rho _q\left( x^N,t_0\right) \right) =0. \label{bc}
\ee
%
%	\
%
Here, the limit $t_0\to -\infty$ is made after thermodynamical
one $N\!\to\infty, V~\to~\infty$, $N/V\to {\rm const}$, $\im=\sqrt{-1}$ and
$L_N$ is the Liouville operator:
%
%	2
%
$$
L_N\ \ = \sum\limits_{j=1}^NL\left( j\right) +\frac 12
\mathop{\sum_{j=1}^N\sum_{k=1}^N}\limits_{j\ne k} L\left(j,k\right),
$$
\be
L\left( j\right) = -\imath \frac{\bmv{p}_j}{2m}\frac \partial {\partial
\bmv{r}_j}, \hspace{1cm}
L\left( j,k\right) = \imath \frac{\partial {\mit \Phi} \left( \left| \bmv{r}
_{jk}\right| \right) }{\partial \bmv{r}_{jk}}\left( \frac \partial {\partial
\bmv{p}_j}-\frac \partial {\partial \bmv{p}_k}\right),	\label{lo}
\ee
%
%	\
%
${\mit \Phi}_{jk}$ is the interaction energy between two particles $j$
and $k$; $x_j=\{\bmv{r},\bmv{p}\}$ is the set of phase variables
(coordinates and momenta). Quasi-equilibrium distribution function
$\rho_q\lp x^N;t\rp$ is determined from the condition of maximum for
informational entropy at fixed values of the single particle distribution
function $f_1\lp x_1;t\rp$ and average density of the interaction energy
$\langle {\cal E}_{int}\lp\bmv{r}\rp\rangle ^t$,
$\displaystyle\lp\langle\ldots\rangle ^t=\int d\Gamma_N\ldots\times\right.$
$\rho\lp x^N;t\rp$, $\left.\displaystyle d\Gamma_N=\frac{\lp dx\rp^N}{N!}\rp$,
that corresponds to taking into account correlations, re\-la\-ted to
con\-ser\-va\-ti\-ons laws of hydrodynamical variables for particle density
$n\lp\bmv{r};t\rp$, momentum $\bmv{j}\lp\bmv{r};t\rp$
and full energy ${\cal E}\lp\bmv{r};t\rp$ \cite{c17}. This function can be
presented as follows \cite{c6,c13}:
%
%	3
%
\be
\rho _q\left( x^N,t\right) =\exp\left( -U_N\left(\bmv{r}^N;t\right)\right)
\prod_{j=1}^N\frac{f_1\left(x_j;t\right)}{u\left(\bmv{r}_j;t\right)},
\label{qso}
\ee
%
%	\
%
where $u\left(\bmv{r}_j;t\right)$ is obtained from the relations:
%
%	n
%
$$
u\left(\bmv{r}_1;t\right) =\int\frac{d\bmv{r}^{N-1}}{\left( N-1\right) !}
\exp\left( -U_N\left(\bmv{r}_1,\bmv{r}^{N-1};t\right)\right)
\prod_{j=2}^N\frac{n\left(\bmv{r}_j;t\right)}{u\left(\bmv{r}_j;t\right)},
$$
%	n
$$
U_N\left(\bmv{r}^N;t\right) =\sum_{j<k}^N {\mit \Phi}_{jk}\beta_{jk},
\hspace{1cm}
\beta_{jk}=\beta\left(\bmv{r}_j,\bmv{r}_k;t\right) \equiv
\frac12\Big(\beta\left(\bmv{r}_j;t\right) +\beta\left(\bmv{r}_k;t\right)
\Big),
$$
%
%	\
%
$\displaystyle n\lp\bmv{r};t\rp=\int d\bmv{p}f_1(x;t)$ is non-equilibrium
particles concentration, $\beta$ is certain function, being an analogue of
local inverse temperature.

Taking into account the boundary condition (\ref{bc}) is equivalent to
the transition from the Liouville equation to a modified one
\cite{c10,c11}:
%
%	4
%
\be
\left( \frac \partial {\partial t}+\imath L_N\right) \rho \left(
x^N,t\right) =-\varepsilon \Big( \rho \left( x^N,t\right) -\rho _q\left(
x^N,t\right) \Big).\label{mle}
\ee
%
%	\
%
This equation contains the small source in the right-hand side, which
destroys the invariance with respect to time inversion ($\varepsilon
\to +0$ after the thermodynamic limit transition). Integrating equation
(\ref{mle}) over the phase space of $(N-s)$ particles, we obtain an equation
chain for the $s$-particle non-equilibrium distribution function
$\displaystyle f_s\lp x^s;t\rp =\int d\Gamma_{N-s}\rho\lp x^N;t\rp$
\cite{c6,c13}:
%
%	5
%
\newcommand{\ds}{\displaystyle}
\bea
&\ds \left( \frac \partial {\partial t}+\imath L_s\right) f_s\left( x^s;t
\right)+\sum\limits_{j=1}^s \int dx_{s+1}\;\imath L\left( j,s+1\right)
f_{s+1}\left(x^{s+1};t\right) = \nonumber \\
\label{mbec}\\
&\ds =-\varepsilon \left( f_s\left( x^s;t\right) -g_s\left(\bmv{r}^s;t\right)
\prod\limits_{j=1}^sf_1\left( x_j;t\right) \right), \nonumber
\eea
%
%	\
%
where
%
%	6
%
\be
g_s\left(\bmv{r}^s;t\right) =\int d\Gamma _{N-s}\; d\bmv{p}^s\;
\rho _q\left( x^N,t\right) \label{qcf}
\ee
%
%	\
%
is the quasi-equilibrium $s$-particle coordinate distribution function
which depends on $n\lp\bmv{r};t\rp$ and $\beta\lp\bmv{r};t\rp$ functionally.
Due to the fact, that $g_1\left(\bmv{r}_1;t\right)=1$, the equation chain
(\ref{mbec}) is distinguished from the ordinary BBGKY hierarchy
\cite{c1} by the availability of sources in the right-hand parts of
the equations beginning from the second one and takes into account both the
one-particle and collective hydrodynamical effects.

Let us consider the solution to equation chain (\ref{mbec}) within the pair
collision approximation. In this case, three- and higher-particle
correlations are neglected, but environment influence to the evolution
of a chosen pair of particles is taken into consideration by means of
application of the boundary condition. Then from (\ref{mbec}) for
$f_2\lp x_1,x_2;t\rp$ we obtain \cite{c6}:
%
%	7
%
\be
\left( \frac \partial {\partial t}+\imath L_2+\varepsilon \right) f_2\left(
x_1,x_2;t\right) =\varepsilon g_2\left(\bmv{r}_1,\bmv{r}_2;t\right)
f_1\left( x_1;t\right) f_1\left( x_2;t\right).	\label{epdf}
\ee
%
%	\
%
The formal solution of equation (\ref{epdf}) is of the form
%
%	8
%
\be
\begin{array}{c}
\ds f_2\left(x_1,x_2;t\right)= \\
\ds =\varepsilon\int\limits_{-\infty}^{0}
d\tau\exp\Big(\lp\varepsilon + \im L_2\rp \tau\Big)
g_2\left(\bmv{r}_1,\bmv{r}_2;t+\tau\right)f_1\left( x_1;t+\tau\right)
f_1\left( x_2;t+\tau\right).
\end{array} \label{pdf}
\ee
%
%	\
%
Substituting (\ref{pdf}) into (\ref{mbec}) at $s=1$ leads to the
kinetic equation for one-particle distribution function in the pair
collision approximation:
%
%	9
%
\be
\lp\frac{\partial}{\partial t} + \im L(1)\rp f_1\lp x_1;t\rp =
I_{col}\lp x_1;t\rp,	\label{ke}
\ee
%
%	\
%
where
%
%	10
%
\be
I_{col}\lp x_1;t\rp =-\int dx_2\im L(1,2)\varepsilon\int\limits_{-\infty}^0
d\tau\exp\Big(\lp\varepsilon + \im L_2\rp \tau\Big)
F_2\left( x_1;t+\tau\right),	\label{ci1}
\ee
%	\
$$
F_2\left( x_1;t+\tau\right)=g_2\left(\bmv{r}_1,\bmv{r}_2;t+\tau\right)
f_1\left( x_1;t+\tau\right)f_1\left( x_2;t+\tau\right)
$$
%
%	\
%
is the collision integral. We must emphasize that equation (\ref{ke}) is
needed to be adjusted with the equation for binary quasi-equilibrium
distribution function $g_2\left(\bmv{r}_1,\bmv{r}_2;t\right)$.
According to (\ref{qso}) and (\ref{qcf}) this function functionally depends
on $n\lp\bmv{r};t\rp$, $\hat{\cal E}_{int}\lp\bmv{r};t\rp$ (or on
$\beta\lp\bmv{r};t\rp$). Besides it was shown, that quasi-equilibrium
correlation distribution function $h_2\lp\bmv{r}_1,\bmv{r}_2;t\rp$, which
is related to $g_2\left(\bmv{r}_1,\bmv{r}_2;t\right)$ \ ($h_2=g_2-1$),
satisfies the Ornstein-Zernike equation \cite{c12}.
In paper \cite{c17a} a non-equilibrium grand canonical distribution for the
system of hard spheres and on the base of methods of non-equilibrium
statistical mechanics \cite{c17b} an Ornstein-Zernike equation for pair
quasi-equilibrium correlation function of hard spheres
$h^{hs}_2(\bmv{r}_1,\bmv{r}_2;t)=g^{hs}_2(\bmv{r}_1,\bmv{r}_2;t)-1$ were
proposed. These results have been generalized in paper \cite{c12}. New
equation for $h_2(\bmv{r}_1,\bmv{r}_2;t)$ is an analogue of
Ornstein-Zernike equation in equilibrium statistical mechanics \cite{c17b}.
This equation for the hard sphere system in equilibrium case has exact
solution in Percus-Yevick approximation \cite{c12}.

In the paper \cite{c6} some special cases were considered, when
the interparticle interaction potential ${\mit\Phi}_{kj}$ is modelled as the
hard sphere potential ${\mit\Phi}^{hs}\lp|\bmv{r}|\rp$ for particles with
diameter $\sigma$. Then taking into account the singularity of the hard
sphere potential ($\tau~\to~+~0$, $\tau$ is collision time) it was shown for
the first time how the collision integral (\ref{ci1}) transforms into
collision integral of revised Enskog theory (RET) \cite{c14}:
%
%	11
%
\be
I_{col}^{hs}\lp x_1;t\rp = \int dx_2\;\hat{T}^{hs}(1,2)
g_2^{hs}\left(\bmv{r}_1,\bmv{r}_2;t\right) f_1\left( x_1;t\right)
f_1\left( x_2;t\right),	\label{hsci1}
\ee
%
%	\
%
where $\hat{T}^{hs}(1,2)$ is the Enskog's collision operator for hard
spheres, $g_2^{hs}\left(\bmv{r}_1,\bmv{r}_2;t\right)$ is the pair
qu\-a\-si\-e\-qui\-lib\-ri\-um distribution function of hard spheres,
which depends on the average non-equilibrium density $n\lp\bmv{r};t\rp$
functionally. It is important to emphasize that $H$-theorem for the kinetic
equation (\ref{ke}) with the Enskog collisions integral (\ref{hsci1}) has
been proved by P.R\'esibois \cite{c15,c16}.

In the same paper \cite{c6} another case was also considered,
when the interparticle interaction potential is modelled as a sum of a
short-range potential (hard spheres, for example) and some long-range
smooth potential;
$$
{\mit \Phi}=\left\{
\begin{array}{ll}
{\mit \Phi}^{hs}, & |\bmv{r}|<\sigma^*; \\
{\mit \Phi}^{l}, & |\bmv{r}|\ge\sigma^*; \\
\end{array}
\right.
$$
where $\sigma^*$ is the effective diameter of hard spheres, which depends
on the method of splitting the potential ${\mit \Phi}(|\bmv{r}|)$
into short- and long-range parts.

If the time retarding and spatial inhomogeneity will be neglected, then
we can present collision integral (\ref{ci1})
in the second approximation with respect to interacting potential
${\mit \Phi}(|\bmv{r}|)$ as follows \cite{c13}:
%
%	12
%
\be
I_{col}\lp x_1;t\rp = I_{col}^{hs}\lp x_1;t\rp +
I_{col}^{mf}\lp x_1;t\rp + I_{col}^{l}\lp x_1;t\rp, \label{ci2}
\ee
%
%	\
%
%
%	13
%
\be
I_{col}^{hs}\lp x_1;t\rp = \int dx_2\;\hat{T}^{hs}(1,2)
g_2\left(\bmv{r}_1,\bmv{r}_2;t\right) f_1\left( x_1;t\right)
f_1\left( x_2;t\right),	\label{hsci2}
\ee
%
%	\
%
%
%	14
%
\be
I_{col}^{mf}\lp x_1;t\rp = \int dx_2\;\imath L^{l}(1,2)
g_2\left(\bmv{r}_1,\bmv{r}_2;t\right) f_1\left( x_1;t\right)
f_1\left( x_2;t\right),	\label{mfci}
\ee
%
%	\
%
%
%	15
%
\bea
\lefteqn{I_{col}^{l}\lp x_1;t\rp =}\label{lci}\\
&&\frac{1}{m}\frac{\partial}{\partial v_{1,\alpha}}
\int d\bmv{g}\; J_{\alpha\beta}(\bmv{g})
\lp \frac{\partial}{\partial v_{1,\beta}}-
\frac{\partial}{\partial v_{2,\beta}} \rp
f_1\left( x_1;t\right)f_1\left( \bmv{r}_1+
\bmv{r}_{12},\bmv{v}_2;t\right),\nonumber
\eea
%
%	\
%
where
%
%	16
%
\bea
\lefteqn{J_{\alpha\beta}(\bmv{g})=}\label{ltci}\\
&&\frac 1m\int\limits_{\sigma^*}^{\infty}
dr_{12}\int d\hat{\sigma}\; r_{12}^2 g_2\left(\bmv{r}_1,
\bmv{r}_1+\bmv{r}_{12};t\right)
\left[\frac{\partial {\mit \Phi}^l(|\bmv{r}_{12}|)}
{\partial r_{12,\alpha}}\right]
\int\limits_{-\infty}^{t}d\tau
\left[\frac{\partial {\mit \Phi}^l(|\bmv{r}_{12}+\bmv{g}\tau|)}
{\partial r_{12,\beta}}
\right],\nonumber
\eea
%
%	\
%
and
$$
\bmv{g}=\bmv{v}_2-\bmv{v}_1,\hspace{2cm}
\hat{\sigma}=(\bmv{r}_2-\bmv{r}_1)/|\bmv{r}_2-\bmv{r}_1|.
$$
The first term in the right hand part of (\ref{ci2}) is the Enskog
collision integral (\ref{hsci2}), where
$g_2\left(\bmv{r}_1,\bmv{r}_2;t\right)$ is the quasi-equilibrium pair
distribution function for system of particles with interaction potential
${\mit\Phi}_{jk}$, which depends on both non-equilibrium concentration
$n\lp\bmv{r};t\rp$ and inverse temperature $\beta\lp\bmv{r};t\rp$
functionally. The second term (\ref{mfci}) is the mean field influence,
and the third one (\ref{lci}) is written in the form of Landau-like
collision integral. If $\displaystyle{\mit\Phi}^l(|\bmv{r}|)=
\frac{(Ze)^2}{r}$ is the
Coulomb potential, equation (\ref{ci2}) with structure of (\ref{ltci})
is a generalization of Landau collision integral without divergency
for short-range distances between particles (that is different from
the usual Landau collision integral \cite{c3,c18}
because short-range interactions are taken into account correctly).
Therefore we can call such kinetic equation (\ref{ke}) with
collision integral (\ref{ci2}) as the Enskog-Landau kinetic equation
for the system of charged hard spheres. It is necessary to emphasize that
in equation (\ref{ltci}) the long-range divergency still remains.
To avoid this problem sequentially we have to consider kinetic equation with
taking into account of dynamical screening effects \cite{c2,c3}. But this way
is impossible in Enskog-Landau kinetic equation. Only one we can do for
further calculation is to change upper integral limit to some finite value,
which could have a meaning of value of statical screening in our system
(see below).
solve this problem we must consider dynamical screening effects.

Following \cite{c13}, we can write for
$I_{col}^{hs}\lp x_1;t\rp$ the next representation:
%
%	17
%
\be
I_{col}^{hs}\lp x_1;t\rp = I_{col}^{hs(0)}\lp x_1;t\rp +
I_{col}^{hs(1)}\lp x_1;t\rp,\label{hscibp}
\ee
%
%	\
%
%
%	18
%
$$
I_{col}^{hs(0)}\lp x_1;t\rp =\int d\bmv{v}_2\; d\varepsilon\;
b\: db\; g\: g_2\left( \sigma ^{+}|n(\bmv{r};t),\beta(\bmv{r};t)
\right) \times
$$
%
%	\
%
\be
\times \Big( f_1\left(\bmv{r}_1,\bmv{v}_1^{\prime };t\right) f_1\left(
\bmv{r}_1,\bmv{v}_2^{\prime };t\right) -f_1\left(\bmv{r}_1,\bmv{v}
_1;t\right) f_1\left(\bmv{r}_1,\bmv{v}_2;t\right) \Big),\label{hscip1}
\ee
%
%	\
%
%
%	19
%
$$
I_{col}^{hs(1)}\left( x_1;t\right) =\sigma ^3\int
d\hat{\bmv{r}}_{12}\; d\bmv{v}_2\; \left(\hat{\bmv{r}}_{12}\bmv{g}\right)
\Theta\left(\hat{\bmv{r}}_{12}\bmv{g}\right)
\hat{\bmv{r}}_{12}g_2\left( \sigma ^{+}|n(\bmv{r};t),
\beta(\bmv{r};t)\right)\times
$$
%
%	\
%
\be
\times \Big( f_1\left(\bmv{r}_1,\bmv{v}_1^{\prime };t\right) \nabla
f_1\left(\bmv{r}_1,\bmv{v}_2^{\prime };t\right) +
f_1\left(\bmv{r}_1,\bmv{v}_1;t\right) \nabla f_1\left(\bmv{r}_1,\bmv{v}
_2;t\right) \Big) \label{hscip2}.
\ee
%
%	\
%
Here we use definitions: $\varepsilon$ is an azimuthal angle of scattering,
$\hat{\bmv{r}}_{12}$ is the unit vector, $\bmv{g}=\bmv{v}_2-\bmv{v}_1$,
$b$ is the impact parameter, $\Theta(x)$ is the unit function,
$\bmv{v}'_1=\bmv{v}_1+\hat{\bmv{r}}_{12}\lp\hat{\bmv{r}}_{12}\cdot\bmv{g}\rp$,
$\bmv{v}'_2=\bmv{v}_2-\hat{\bmv{r}}_{12}\lp\hat{\bmv{r}}_{12}\cdot\bmv{g}\rp$
are velocities of particles after a collision, and $\sigma^+$ is a diameter
of the particle.

By representing (\ref{lci}) in the cylindrical coordinates, one can introduce
the impact parameter $b$, azimuthal angle of scattering $\varepsilon$,
distance along the cylinder axis $\xi$. Then Boltzmann-like collision integral
can be obtained from (\ref{lci}), putting $g_2\equiv 1$:
%
%	20
%
\be
I_{col}^{l}\lp x_1;t\rp =\int d\bmv{v}_2\; d\varepsilon\;
b\: db\; g\Big( f_1\left(\bmv{r}_1,\bmv{v}_1^{*};t\right) f_1\left(
\bmv{r}_1,\bmv{v}_2^{*};t\right) -f_1\left(\bmv{r}_1,\bmv{v}
_1;t\right) f_1\left(\bmv{r}_1,\bmv{v}_2;t\right) \Big),\label{blci}
\ee
%
%	\
%
where
%
%	21
%
\be
\begin{array}{c}
\bmv{v}_1^*=\bmv{v}_1+\Delta\bmv{v},\qquad
\bmv{v}_2^*=\bmv{v}_2-\Delta\bmv{v}, \\
\\
\ds \Delta\bmv{v} = -\frac{1}{mg}\int\limits_{-\infty}^{+\infty}
d\xi\nabla\Phi^l(|\bmv{r}_{12}|)\bigg | _{r_{12}=\sqrt{b^2+\xi^2}}.
\end{array} \label{vc}
\ee
%
%	\
%

After these transformations we have the kinetic equation (\ref{ke}), where
the collision integral $I_{col}(x_1;t)$ can be presented as the sum of
(\ref{mfci}), (\ref{hscip1}), (\ref{hscip2}) and (\ref{blci}).
\vspace{5ex}

\section{Normal solution. Transport coefficients}

     We shall solve the equation (\ref{ke}) by means of  iterations.
Therefore  a set of hydrodynamical variables should be introduced for
reduced description of the system: hydrodynamical density of mass, momentum
and kinetic energy \cite{c4,c19}. The conservation laws are to be written
down after multiplication of both left and right parts of equation
(\ref{ke}) by components of additive invariant vector $m$, $m\bmv{v}$
as well as by kinetic energy $\displaystyle \frac{mc^2}{2}$ and after
integrating over $\bmv{v}$  \cite{c4,c19}. The local-equilibrium Maxwell
distribution function may  be taken as an initial approximation:
%
%	22
%
\be
f_1^{\left( 0\right) }\left( x_1;t\right) =n\left(\bmv{r}_1;t\right) \left(
\frac m{2\pi kT\left(\bmv{r}_1;t\right) }\right) ^{3/2}\exp \left( -
\frac{mc_1^2\left(\bmv{r}_1;t\right) }{2kT\left(\bmv{r}_1;t\right) }
\right),\label{mdf}
\ee
%
%	\
%
where $n(\bmv{r};t)$ is the density,
$\bmv{c}(\bmv{r};t) = \bmv{v} -\bmv{V}(\bmv{r};t)$, and
$\bmv{V}(\bmv{r};t)$ is the hydrodynamical velocity. The total distribution
function  $f_1(x;t)$ has a form $f^0_1(x;t)\Big( 1+\varphi(x_1;t)\Big)$ and
the correction $\varphi(x_1;t)$ may be expressed through Sonine-Laguerre
polynomials \cite{c4}. The calculations show \cite{c13}:
%
%	23
%
$$
\varphi \left( x_1;t\right) =\frac{A\left( 1\right) }{T\left(\bmv{r}
_1;t\right) }\sqrt{\frac m{2kT\left(\bmv{r}_1;t\right) }}\left(
\frac 52-\frac{mc_1^2\left(\bmv{r}_1;t\right) }{2kT\left(\bmv{r}
_1;t\right) }\right) \left(\bmv{c}_1\cdot \nabla \right) T\left(\bmv{r}
_1;t\right) -
$$
%
%	\
%
\be
-\frac{mB\left( 0\right) }{2kT\left(\bmv{r}_1;t\right) }\left(\bmv{c}_1
\bmv{c}_1-\frac 13c_1^2\stackrel{\leftrightarrow }{I}\right) :\left(
\nabla V\left(\bmv{r}_1;t\right) \right)\label{correction}
\ee
%
%	\
%
where $\stackrel{\leftrightarrow }{I}$ is the unit tensor;
$A(1)$, $B(0)$ are coefficients which satisfy the following relations:
%
%	24
%
\be
A\left( 1\right) =\frac{15}{8}\sqrt{\frac{\pi}{2}}\;\;\,
\times\frac{1+\frac 25\pi
n\sigma^3g_2\left(\sigma^{+}|n,\beta\right)}
{n\Big( g_2\left(\sigma^{+}|n,\beta\right)
\Omega _{hs}^{\left( 2,2\right) }+\Omega _l^{\left( 2,2\right)
}\Big) },\label{a1}
\ee
%
%	\
%
%
%	25
%
\be
B\left( 0\right) =\frac{5}{2}\sqrt{\frac{\pi m}{kT}}
\times\frac{1+\frac 4{15}\pi n\sigma^3
g_2\left(\sigma^{+}|n,\beta\right)}
{n\Big( g_2\left(\sigma^{+}|n,\beta\right)
\Omega_{hs}^{\left( 2,2\right) }+
\Omega _l^{\left( 2,2\right) }\Big)}.\label{b0}
\ee
%
%	\
%
Here, we use notations:
%
%	26
%
\be
\Omega _{hs,l}^{\left( p,q\right) }=\int\limits_0^\infty dg_0g^{2q+3}_0\exp
\left( -g^2_0\right) \Omega _{hs,l}^{\left( p\right) },\label{oi}
\ee
%
%	\
%
%
%	27
%
\be
\Omega _{hs}^{\left( p\right) }=2\pi \int\limits_0^\sigma bdb\Big( 1-\cos
{}^p\chi ^{\prime }\left( b,g\right) \Big) ,\label{hsoi}
\ee
%
%	\
%
%
%	28
%
\be
\Omega _l^{\left( p\right) }=2\pi \int\limits_\sigma ^\infty bdb\Big(
1-\cos {}^p\chi ^{*}\left( b,g\right) \Big) , \label{coi}
\ee
%
%	\
%
%
%
%
$$
\bmv{g}_0=\sqrt{\frac{m}{2kT}}\,\bmv{g}.
$$
The expressions (\ref{oi}-\ref{coi}) are known as $\Omega$-integrals
\cite{c4}, $\chi'$, $\chi^*$ being the angles of scattering for the
hard spheres and Coulomb particles respectively.

     The $\Omega$-integrals can be calculated exactly \cite{c4} or
approximately, if the first way fails. We shall attempt to do this from the
geometrical point of view. Considering the dynamics of collision of hard
spheres, we have \cite{c4}
%
%	29
%
\be
\cos \frac{\chi '}{2}=\frac{b}{2}, \hspace{1cm}{\rm and}\hspace{1cm}
\Omega_{hs}^{(2,2)}=2\pi\sigma^2.\label{hsoi1}
\ee
%
%	\
%
If the angles of scattered charged particles are  assumed  to
be small, the following approximation can be obtained \cite{c19a}:
%
%	30
%
\be
\sin \chi^*\approx 2\frac{(Ze)^2}{mg^2_0}\int\limits_0^\infty
\frac{d\xi}{\lp b^2+\xi^2\rp ^{\frac 32}}.\label{sinchi}
\ee
%
%	\
%
This expression for $\sin \chi^*$  leads to logarithmical divergency
in $\Omega_l^{(2,2)}$ at integration over infinite sight parameter
(\ref{coi}). To avoid this difficulty, the infinite sight parameter in
the integral (\ref{coi}) for  calculating $\Omega_l^{(2,2)}$ should  be
replaced by the Debye-type finite radius $D$ of screening for such
system. Then we obtain:
%
%	31
%
\be
\Omega_l^{(2,2)}=\pi^3\frac{(Ze)^4}{( kT)^2}
\ln\frac D\sigma.\label{coi1}
\ee
%
%	\
%

     The stress tensor and heat flux vector for obtained distribution
function $f_1(x_1;t)$ in the first approximation are:
%
%	32
%
\be
\stackrel{\leftrightarrow}{P}\left( \bmv{r}_1;t\right) =P\left(
\bmv{r}_1;t\right)
\stackrel{\leftrightarrow }{I}-\ae\Big(\nabla
V\left(\bmv{r}_1;t\right)\Big)-2\eta
\stackrel{\leftrightarrow}{S}\left(\bmv{r}_1;t\right),\label{st}
\ee
%
%	\
%
where $P$ is the pressure, $\stackrel{\leftrightarrow}{S}(\bmv{r}_1;t)$ is
the shift tensor of velocities, $\ae$ is the coefficient of bulk viscosity
%
%	33
%
\be
\ae =\frac 49\sigma ^4n^2
g_2\left( \sigma ^{+}|n,\beta\right)\sqrt{\pi mkT},\label{bv}
\ee
%
%	\
%
$\eta$ is the coefficient of shear viscosity
%
%	34
%
\be
\eta =\frac 35\ae +
\frac 12nkT
\left( 1+\frac 4{15}\pi n\sigma ^3g_2\left(\sigma^{+}|n,\beta
\right)\right) B\left( 0\right).\label{sv}
\ee
%
%	\
%
The heat flux vector has the form
%
%	35
%
\be
\bmv{q}(\bmv{r}_1;t) =
-\lambda\left(\nabla\cdot T(\bmv{r}_1;t)\right),\label{hfv}
\ee
%
%	\
%
where $\lambda$ is the thermal conductivity:
%
%	36
%
\be
\lambda =\frac{3k}{2m}\ae +\frac 54nk\sqrt{\frac{2kT}m}
\left( 1+\frac
25\pi n\sigma^3g_2\left(\sigma^{+}|n,\beta\right)\right)
A\left( 1\right).\label{tc}
\ee
%
%	\
%
We can consider some particular expressions for the quantities $\ae$,
$\eta$ and $\lambda$. If $n\to 0$, then $\ae$, $\eta$, and $\lambda$
coincide with respective quantities obtained by solving Boltzmann's equation
for low density systems  of point-like charged particles \cite{c19}.
If $Z=0$,  then  we  obtain the results of RET-theory \cite{c14}.

\section{Numerical calculations}

The numerical calculations were carried out for the viscosities $\ae$
(\ref{bv}), $\eta$ (\ref{sv}) and thermal conductivity $\lambda$ (\ref{tc}),
where the dense once-ionized argon was chosen as a dense one-component
plasma in a homogeneous neutralizing continuum. In the case $Z=0$, the
obtained results were compared with \cite{c20}. In these papers dense
neutral argon was considered. Its atoms were modelled as hard spheres,
and obtained results are in a good agreement with \cite{c20} if the theory
parameter $\sigma$ is chosen correctly.

The binary correlation distribution function $g_2(\sigma^{+}|n,\beta)$
was taken from \cite{c21} where it is presented as the functional
of $\sigma$ and $n$, but not of $\beta$:
%
%	37
%
\be
g_2\Big(\sigma,n\Big)=
\lp 1-\frac{\pi}{12}n\sigma^3\rp\times
\lp 1-\frac{\pi}{6}n\sigma^3\rp^{-3}.
\ee
%
%	\
%
The screening radius $D$ was not chosen as a Debye-like one, because
such approximation can give for high densities incorrect values being less
than $\sigma$. Therefore
it is presented in the form, that was proposed in \cite{c22}, where the
hard spheres diameter is also taken into account:
%
%	38
%
\be
D=\sigma \frac{1-\Delta}{1+2\Delta}\left[\sqrt{1+4Ze\sigma
\lp\frac{1-\Delta}{1+2\Delta}\rp ^2
\sqrt{\frac{\pi n}{kT}}\ }
-1\right]^{-1},
\ee
%
%	\
%
where
$$
\Delta=\frac{1}{6}\pi n\sigma^3.
$$
Such substitution of $\sigma$, $D$ and $g_2(\sigma,n)$ allows to obtain
transport coefficients as  functions of density $n$ and temperature $T$.
The behaviours of $\ae$, $\eta$ and $\lambda$ were investigated for various
values of $n$ and $T$ in\-clu\-ding regions mentioned in \cite{c20},
but only for the case giving value of $D$ greater than $\sigma$. The purpose
of these calculations was to study the dependence of $\ae$, $\eta$ and
$\lambda$ on the long-range potential when the density is high. As was
expected the behaviour of such transport coefficients in these regions
appears to be rather smooth and monotonous. But the small deviation is
observed between our results and comparative ones when the long
interaction is ``switched on''. Transport coefficients appear to be
slightly sensitive to including the long-range potential. They decrease
slowly being of the same order. These behaviour coincide with expectations.

We note that the numerical calculation for $\ae$, $\eta$ and $\lambda$ was
carried out and compared with data from \cite{c23,c24,c25}. As one can see
from the figures below, the good coincidence of both data is observed in
the ``high''-temperature region. Theory parameter $\sigma$ for different
densities was borrowed from \cite{c26,c27,c28,c29}.
It is possible to improve the results by choosing $\sigma$ more precise but
in general case $\sigma$ is the function of $n$, $T$, $Z$ and finding
for this dependence is a microscopic problem.

As it can be seen from the figures below, the most essential deviations
between of theoretical calculations from the experimental data arise in the
low-temperature region for
$\eta$ (figure~\ref{f1}) and for $\lambda$ (figure~\ref{f2}).
In real physical systems at low temperature one can approach to gas-liquid
phase transition point. But our theory is not applicable within critical
region near point of phase transition. Except that there is no phase
transition in one-component system with only repulsive potential on the
contrary to real systems. So our theory works well for description of
transport processes at high temperatures far from phase transition point
of corresponding real systems.

%These deviations increase approaching the gas-liquid phase transition
%point, just as in critical region. It is well-known that taking into account
%both li\-ne\-ar and nonlinear fluctuations is of great importance for
%description of system behaviour in critical region.
%No fluctuations are considered in present theory. Therefore it
%can be used only far from phase transition point, that is confirmed by all
%illustrations.

\section*{Acknowledgements}

This work was supported partially by the State Fund for Fundamental
Investigations at Ukrainian State Committee for Sciences and Technology,
Project No~2.3/371.

%%%%%%%%%%%%%%%%%%%%%%%%%%%%%%%%%%%%%%%%%%%%%%%%%%%%%%%%%%%%%%%%%%%%%%%%%%%%%
%
%	R E F E R E N C E S
%
%%%%%%%%%%%%%%%%%%%%%%%%%%%%%%%%%%%%%%%%%%%%%%%%%%%%%%%%%%%%%%%%%%%%%%%%%%%%%

%%%%%%%%%%%%%%%%%%%%%%%%%%%%%%%%%%%%%%%%%%%%%%%%%%%%%%%%%%%%%%%%%%%%%%%%%%%%%
%                                                                           %
%	F I G U R E S                                                       %
%                                                                           %
%%%%%%%%%%%%%%%%%%%%%%%%%%%%%%%%%%%%%%%%%%%%%%%%%%%%%%%%%%%%%%%%%%%%%%%%%%%%%

\newpage
\begin{figure}[ht]
\unitlength=1in
\special{em:linewidth 0.4pt}
\linethickness{0.4pt}
\def\emmv{\special{em:moveto}}
\def\emln{\special{em:lineto}}
\def\tmls{\normalsize}
\begin{picture}(5.6025,3.531)(1.5975,1.619)
\special{em:linewidth 0.4pt}
\linethickness{0.4pt}
\put(2.2,1.9){\emmv}
\put(7.2,1.9){\emln}
\put(2.2,1.9){\emmv}
\put(2.2,1.844){\emln}
\put(2.1375,1.619){\makebox(0,0)[lb]{\tmls 4}}
\put(2.7,1.9){\emmv}
\put(2.7,1.8625){\emln}
\put(3.2,1.9){\emmv}
\put(3.2,1.844){\emln}
\put(3.1375,1.619){\makebox(0,0)[lb]{\tmls 6}}
\put(3.7,1.9){\emmv}
\put(3.7,1.8625){\emln}
\put(4.2,1.9){\emmv}
\put(4.2,1.844){\emln}
\put(4.1375,1.619){\makebox(0,0)[lb]{\tmls 8}}
\put(4.7,1.9){\emmv}
\put(4.7,1.8625){\emln}
\put(5.2,1.9){\emmv}
\put(5.2,1.844){\emln}
\put(5.075,1.619){\makebox(0,0)[lb]{\tmls 10}}
\put(5.7,1.9){\emmv}
\put(5.7,1.8625){\emln}
\put(6.2,1.9){\emmv}
\put(6.2,1.844){\emln}
\put(6.075,1.619){\makebox(0,0)[lb]{\tmls 12}}
\put(6.7,1.9){\emmv}
\put(6.7,1.8625){\emln}
\put(7.2,1.9){\emmv}
\put(7.2,1.844){\emln}
\put(7.075,1.619){\makebox(0,0)[lb]{\tmls 14}}
\special{em:linewidth 0.4pt}
\linethickness{0.4pt}
\put(2.2,1.9){\emmv}
\put(2.2,5.15){\emln}
\put(2.2,1.9){\emmv}
\put(2.1435,1.9){\emln}
\put(1.944,1.8375){\makebox(0,0)[lb]{\tmls 1}}
\put(2.2,2.225){\emmv}
\put(2.1625,2.225){\emln}
\put(2.2,2.55){\emmv}
\put(2.1435,2.55){\emln}
\put(1.944,2.4875){\makebox(0,0)[lb]{\tmls 3}}
\put(2.2,2.875){\emmv}
\put(2.1625,2.875){\emln}
\put(2.2,3.2){\emmv}
\put(2.1435,3.2){\emln}
\put(1.944,3.1375){\makebox(0,0)[lb]{\tmls 5}}
\put(2.2,3.525){\emmv}
\put(2.1625,3.525){\emln}
\put(2.2,3.85){\emmv}
\put(2.1435,3.85){\emln}
\put(1.944,3.7875){\makebox(0,0)[lb]{\tmls 7}}
\put(2.2,4.175){\emmv}
\put(2.1625,4.175){\emln}
\put(2.2,4.5){\emmv}
\put(2.1435,4.5){\emln}
\put(1.944,4.4375){\makebox(0,0)[lb]{\tmls 9}}
\put(2.2,4.825){\emmv}
\put(2.1625,4.825){\emln}
\put(2.2,5.15){\emmv}
\put(2.1435,5.15){\emln}
\put(1.819,5.0875){\makebox(0,0)[lb]{\tmls 11}}
\special{em:linewidth 0.4pt}
\linethickness{0.4pt}
\put(2.7,3.0055){\emmv}
\put(2.75,3.0195){\emln}
\put(2.8,3.0335){\emln}
\put(2.85,3.0475){\emln}
\put(2.9,3.0615){\emln}
\put(2.95,3.075){\emln}
\put(3,3.0885){\emln}
\put(3.05,3.102){\emln}
\put(3.1,3.1155){\emln}
\put(3.15,3.129){\emln}
\put(3.2,3.142){\emln}
\put(3.25,3.155){\emln}
\put(3.3,3.168){\emln}
\put(3.35,3.1805){\emln}
\put(3.4,3.193){\emln}
\put(3.45,3.206){\emln}
\put(3.5,3.2185){\emln}
\put(3.55,3.2305){\emln}
\put(3.6,3.243){\emln}
\put(3.65,3.2555){\emln}
\put(3.7,3.2675){\emln}
\put(3.75,3.2795){\emln}
\put(3.8,3.2915){\emln}
\put(3.85,3.3035){\emln}
\put(3.9,3.315){\emln}
\put(3.95,3.327){\emln}
\put(4,3.3385){\emln}
\put(4.05,3.35){\emln}
\put(4.1,3.3615){\emln}
\put(4.15,3.373){\emln}
\put(4.2,3.3845){\emln}
\put(4.25,3.3955){\emln}
\put(4.3,3.4065){\emln}
\put(4.35,3.418){\emln}
\put(4.4,3.429){\emln}
\put(4.45,3.44){\emln}
\put(4.5,3.451){\emln}
\put(4.55,3.462){\emln}
\put(4.6,3.4725){\emln}
\put(4.65,3.4835){\emln}
\put(4.7,3.494){\emln}
\put(4.75,3.5045){\emln}
\put(4.8,3.515){\emln}
\put(4.85,3.5255){\emln}
\put(4.9,3.536){\emln}
\put(4.95,3.5465){\emln}
\put(5,3.557){\emln}
\put(5.05,3.567){\emln}
\put(5.1,3.5775){\emln}
\put(5.15,3.5875){\emln}
\put(5.2,3.598){\emln}
\put(5.25,3.608){\emln}
\put(5.3,3.618){\emln}
\put(5.35,3.628){\emln}
\put(5.4,3.638){\emln}
\put(5.45,3.648){\emln}
\put(5.5,3.6575){\emln}
\put(5.55,3.6675){\emln}
\put(5.6,3.677){\emln}
\put(5.65,3.687){\emln}
\put(5.7,3.6965){\emln}
\put(5.75,3.706){\emln}
\put(5.8,3.7155){\emln}
\put(5.85,3.7255){\emln}
\put(5.9,3.735){\emln}
\put(5.95,3.744){\emln}
\put(6,3.7535){\emln}
\put(6.05,3.763){\emln}
\put(6.1,3.7725){\emln}
\put(6.15,3.7815){\emln}
\put(6.2,3.791){\emln}
\put(6.25,3.8){\emln}
\put(6.3,3.8095){\emln}
\put(6.35,3.8185){\emln}
\put(6.4,3.8275){\emln}
\put(6.45,3.8365){\emln}
\put(6.5,3.8455){\emln}
\put(6.55,3.8545){\emln}
\put(6.6,3.8635){\emln}
\put(6.65,3.8725){\emln}
\special{em:linewidth 0.4pt}
\linethickness{0.4pt}
\put(2.7,3.292){\emmv}
\put(2.95,3.2915){\emln}
\put(3.2,3.3065){\emln}
\put(3.45,3.3285){\emln}
\put(3.7,3.358){\emln}
\put(3.95,3.3925){\emln}
\put(4.2,3.431){\emln}
\put(4.45,3.4715){\emln}
\put(4.7,3.5145){\emln}
\put(5.2,3.603){\emln}
\put(5.7,3.6945){\emln}
\put(6.2,3.787){\emln}
\put(6.7,3.88){\emln}
\special{em:linewidth 0.4pt}
\linethickness{0.4pt}
\put(3.2,3.3065){\makebox(0,0)[cc]{$\ast$}}
\put(4.2,3.431){\makebox(0,0)[cc]{$\ast$}}
\put(5.7,3.6945){\makebox(0,0)[cc]{$\ast$}}
\special{em:linewidth 0.4pt}
\linethickness{0.4pt}
\put(4.4765,4.6345){\makebox(0,0)[lb]{\tmls {\Huge Ar}}}
\special{em:linewidth 0.4pt}
\linethickness{0.4pt}
\put(6.785,2.0755){\makebox(0,0)[lb]{\tmls {\Large T}}}
\special{em:linewidth 0.4pt}
\linethickness{0.4pt}
\put(1.5975,4.7765){\Large\makebox(0,0)[cc]{$\eta$}}
\special{em:linewidth 0.4pt}
\linethickness{0.4pt}
\put(2.2,5.15){\emmv}
\put(7.2,5.15){\emln}
\special{em:linewidth 0.4pt}
\linethickness{0.4pt}
\put(7.2,1.9){\emmv}
\put(7.2,5.15){\emln}
\end{picture}
\caption{Temperature dependence of shear viscosity $\eta$ of neutral {\sf Ar} 
at $\Delta=0.1$ ($n=4.86\cdot10^{21}$ cm$^{-3}$). Solid line represents 
results from theory, solid marked line represents data of 
\protect\cite{c23,c24}. Both $\eta$ and $T$ are dimensionless. The 
transition relations to dimensional data read: $\eta_{dim}=\eta\cdot10^{-5}\ 
Pa\cdot sec.$, $T_{dim}=T\cdot10^{2}\ K$.}
\label{f1}
\vspace*{5ex}
%\end{figure}
%\newpage
%
%
%
%\begin{figure}[h]
\unitlength=1in
\special{em:linewidth 0.4pt}
\linethickness{0.4pt}
\def\emmv{\special{em:moveto}}
\def\emln{\special{em:lineto}}
\def\tmls{\normalsize}
\begin{picture}(5.6025,3.531)(1.5975,1.619)
\special{em:linewidth 0.4pt}
\linethickness{0.4pt}
\put(2.2,1.9){\emmv}
\put(7.2,1.9){\emln}
\put(2.2,1.9){\emmv}
\put(2.2,1.844){\emln}
\put(2.1375,1.619){\makebox(0,0)[lb]{\tmls 3}}
\put(2.5845,1.9){\emmv}
\put(2.5845,1.8625){\emln}
\put(2.969,1.9){\emmv}
\put(2.969,1.844){\emln}
\put(2.9065,1.619){\makebox(0,0)[lb]{\tmls 5}}
\put(3.354,1.9){\emmv}
\put(3.354,1.8625){\emln}
\put(3.7385,1.9){\emmv}
\put(3.7385,1.844){\emln}
\put(3.676,1.619){\makebox(0,0)[lb]{\tmls 7}}
\put(4.123,1.9){\emmv}
\put(4.123,1.8625){\emln}
\put(4.5075,1.9){\emmv}
\put(4.5075,1.844){\emln}
\put(4.445,1.619){\makebox(0,0)[lb]{\tmls 9}}
\put(4.8925,1.9){\emmv}
\put(4.8925,1.8625){\emln}
\put(5.277,1.9){\emmv}
\put(5.277,1.844){\emln}
\put(5.152,1.619){\makebox(0,0)[lb]{\tmls 11}}
\put(5.6615,1.9){\emmv}
\put(5.6615,1.8625){\emln}
\put(6.046,1.9){\emmv}
\put(6.046,1.844){\emln}
\put(5.921,1.619){\makebox(0,0)[lb]{\tmls 13}}
\put(6.431,1.9){\emmv}
\put(6.431,1.8625){\emln}
\put(6.8155,1.9){\emmv}
\put(6.8155,1.844){\emln}
\put(6.6905,1.619){\makebox(0,0)[lb]{\tmls 15}}
\put(7.2,1.9){\emmv}
\put(7.2,1.8625){\emln}
\special{em:linewidth 0.4pt}
\linethickness{0.4pt}
\put(2.2,1.9){\emmv}
\put(2.2,5.15){\emln}
\put(2.2,1.9){\emmv}
\put(2.1435,1.9){\emln}
\put(1.944,1.8375){\makebox(0,0)[lb]{\tmls 0}}
\put(2.2,2.225){\emmv}
\put(2.1625,2.225){\emln}
\put(2.2,2.55){\emmv}
\put(2.1435,2.55){\emln}
\put(1.944,2.4875){\makebox(0,0)[lb]{\tmls 3}}
\put(2.2,2.875){\emmv}
\put(2.1625,2.875){\emln}
\put(2.2,3.2){\emmv}
\put(2.1435,3.2){\emln}
\put(1.944,3.1375){\makebox(0,0)[lb]{\tmls 6}}
\put(2.2,3.525){\emmv}
\put(2.1625,3.525){\emln}
\put(2.2,3.85){\emmv}
\put(2.1435,3.85){\emln}
\put(1.944,3.7875){\makebox(0,0)[lb]{\tmls 9}}
\put(2.2,4.175){\emmv}
\put(2.1625,4.175){\emln}
\put(2.2,4.5){\emmv}
\put(2.1435,4.5){\emln}
\put(1.819,4.4375){\makebox(0,0)[lb]{\tmls 12}}
\put(2.2,4.825){\emmv}
\put(2.1625,4.825){\emln}
\put(2.2,5.15){\emmv}
\put(2.1435,5.15){\emln}
\put(1.819,5.0875){\makebox(0,0)[lb]{\tmls 15}}
\special{em:linewidth 0.4pt}
\linethickness{0.4pt}
\put(2.5845,2.5845){\emmv}
\put(2.623,2.593){\emln}
\put(2.6615,2.6015){\emln}
\put(2.7,2.61){\emln}
\put(2.7385,2.618){\emln}
\put(2.777,2.626){\emln}
\put(2.8155,2.634){\emln}
\put(2.854,2.642){\emln}
\put(2.8925,2.65){\emln}
\put(2.931,2.6575){\emln}
\put(2.969,2.6655){\emln}
\put(3.0075,2.673){\emln}
\put(3.046,2.6805){\emln}
\put(3.0845,2.688){\emln}
\put(3.123,2.6955){\emln}
\put(3.1615,2.7025){\emln}
\put(3.2,2.71){\emln}
\put(3.2385,2.717){\emln}
\put(3.277,2.7245){\emln}
\put(3.3155,2.7315){\emln}
\put(3.354,2.7385){\emln}
\put(3.3925,2.7455){\emln}
\put(3.431,2.7525){\emln}
\put(3.469,2.759){\emln}
\put(3.5075,2.766){\emln}
\put(3.546,2.7725){\emln}
\put(3.5845,2.7795){\emln}
\put(3.623,2.786){\emln}
\put(3.6615,2.7925){\emln}
\put(3.7,2.799){\emln}
\put(3.7385,2.8055){\emln}
\put(3.777,2.812){\emln}
\put(3.8155,2.8185){\emln}
\put(3.854,2.825){\emln}
\put(3.8925,2.831){\emln}
\put(3.931,2.8375){\emln}
\put(3.969,2.8435){\emln}
\put(4.0075,2.85){\emln}
\put(4.046,2.856){\emln}
\put(4.0845,2.862){\emln}
\put(4.123,2.868){\emln}
\put(4.1615,2.874){\emln}
\put(4.2,2.88){\emln}
\put(4.2385,2.886){\emln}
\put(4.277,2.892){\emln}
\put(4.3155,2.898){\emln}
\put(4.354,2.904){\emln}
\put(4.3925,2.9095){\emln}
\put(4.431,2.9155){\emln}
\put(4.469,2.921){\emln}
\put(4.5075,2.927){\emln}
\put(4.546,2.9325){\emln}
\put(4.5845,2.938){\emln}
\put(4.623,2.944){\emln}
\put(4.6615,2.9495){\emln}
\put(4.7,2.955){\emln}
\put(4.7385,2.9605){\emln}
\put(4.777,2.966){\emln}
\put(4.8155,2.9715){\emln}
\put(4.854,2.977){\emln}
\put(4.8925,2.9825){\emln}
\put(4.931,2.988){\emln}
\put(4.969,2.993){\emln}
\put(5.0075,2.9985){\emln}
\put(5.046,3.004){\emln}
\put(5.0845,3.009){\emln}
\put(5.123,3.0145){\emln}
\put(5.1615,3.0195){\emln}
\put(5.2,3.025){\emln}
\put(5.2385,3.03){\emln}
\put(5.277,3.035){\emln}
\put(5.3155,3.0405){\emln}
\put(5.354,3.0455){\emln}
\put(5.3925,3.0505){\emln}
\put(5.431,3.0555){\emln}
\put(5.469,3.0605){\emln}
\put(5.5075,3.066){\emln}
\put(5.546,3.071){\emln}
\put(5.5845,3.076){\emln}
\put(5.623,3.0805){\emln}
\put(5.6615,3.0855){\emln}
\put(5.7,3.0905){\emln}
\put(5.7385,3.0955){\emln}
\put(5.777,3.1005){\emln}
\put(5.8155,3.1055){\emln}
\put(5.854,3.11){\emln}
\put(5.8925,3.115){\emln}
\put(5.931,3.12){\emln}
\put(5.969,3.1245){\emln}
\put(6.0075,3.1295){\emln}
\put(6.046,3.134){\emln}
\put(6.0845,3.139){\emln}
\put(6.123,3.1435){\emln}
\put(6.1615,3.1485){\emln}
\put(6.2,3.153){\emln}
\put(6.2385,3.1575){\emln}
\put(6.277,3.1625){\emln}
\put(6.3155,3.167){\emln}
\put(6.354,3.1715){\emln}
\put(6.3925,3.176){\emln}
\special{em:linewidth 0.4pt}
\linethickness{0.4pt}
\put(2.5845,2.875){\emmv}
\put(2.777,2.8535){\emln}
\put(2.969,2.8445){\emln}
\put(3.1615,2.8425){\emln}
\put(3.354,2.8445){\emln}
\put(3.546,2.847){\emln}
\put(3.7385,2.8795){\emln}
\put(3.931,2.8925){\emln}
\put(4.123,2.9095){\emln}
\put(4.3155,2.938){\emln}
\put(4.5075,2.9595){\emln}
\put(4.7,2.981){\emln}
\put(4.8925,3.005){\emln}
\put(5.277,3.0395){\emln}
\put(5.6615,3.0745){\emln}
\put(6.046,3.1305){\emln}
\put(6.431,3.148){\emln}
\special{em:linewidth 0.4pt}
\linethickness{0.4pt}
\put(2.969,2.8445){\makebox(0,0)[cc]{$\ast$}}
\put(3.7385,2.8795){\makebox(0,0)[cc]{$\ast$}}
\put(4.5075,2.9595){\makebox(0,0)[cc]{$\ast$}}
\put(5.6615,3.0745){\makebox(0,0)[cc]{$\ast$}}
\special{em:linewidth 0.4pt}
\linethickness{0.4pt}
\put(4.4765,4.6345){\makebox(0,0)[lb]{\tmls {\Huge Ar}}}
\special{em:linewidth 0.4pt}
\linethickness{0.4pt}
\put(6.785,2.0755){\makebox(0,0)[lb]{\tmls {\Large T}}}
\special{em:linewidth 0.4pt}
\linethickness{0.4pt}
\put(1.5975,4.7765){\Large\makebox(0,0)[cc]{$\lambda$}}
\special{em:linewidth 0.4pt}
\linethickness{0.4pt}
\put(2.2,5.15){\emmv}
\put(7.2,5.15){\emln}
\special{em:linewidth 0.4pt}
\linethickness{0.4pt}
\put(7.2,1.9){\emmv}
\put(7.2,5.15){\emln}
\end{picture}
\caption{Temperature dependence of thermal conductivity $\lambda$ of neutral
{\sf Ar} at $\Delta=0.075$ ($n=3.644\cdot10^{21}$ cm$^{-3}$). The legend is 
the same as for figure \protect\ref{f1}. Both $\lambda$ and $T$ are
dimensionless. The transition relations to dimensional data read: 
$\lambda_{dim}=\lambda\cdot10^{-2}\ Wt/(m\cdot K)$, $T_{dim}=T\cdot10^{2}\ 
K$.}
\label{f2}
\end{figure}
\clearpage
\begin{figure}[ht]
\unitlength=1in
\special{em:linewidth 0.4pt}
\linethickness{0.4pt}
\def\emmv{\special{em:moveto}}
\def\emln{\special{em:lineto}}
\def\tmls{\normalsize}
\begin{picture}(5.6025,3.531)(1.5975,1.619)
\special{em:linewidth 0.4pt}
\linethickness{0.4pt}
\put(2.2,1.9){\emmv}
\put(7.2,1.9){\emln}
\put(2.2,1.9){\emmv}
\put(2.2,1.844){\emln}
\put(2.1375,1.619){\makebox(0,0)[lb]{\tmls 0}}
\put(2.557,1.9){\emmv}
\put(2.557,1.8625){\emln}
\put(2.9145,1.9){\emmv}
\put(2.9145,1.844){\emln}
\put(2.852,1.619){\makebox(0,0)[lb]{\tmls 8}}
\put(3.2715,1.9){\emmv}
\put(3.2715,1.8625){\emln}
\put(3.6285,1.9){\emmv}
\put(3.6285,1.844){\emln}
\put(3.5035,1.619){\makebox(0,0)[lb]{\tmls 16}}
\put(3.9855,1.9){\emmv}
\put(3.9855,1.8625){\emln}
\put(4.343,1.9){\emmv}
\put(4.343,1.844){\emln}
\put(4.218,1.619){\makebox(0,0)[lb]{\tmls 24}}
\put(4.7,1.9){\emmv}
\put(4.7,1.8625){\emln}
\put(5.057,1.9){\emmv}
\put(5.057,1.844){\emln}
\put(4.932,1.619){\makebox(0,0)[lb]{\tmls 32}}
\put(5.4145,1.9){\emmv}
\put(5.4145,1.8625){\emln}
\put(5.7715,1.9){\emmv}
\put(5.7715,1.844){\emln}
\put(5.6465,1.619){\makebox(0,0)[lb]{\tmls 40}}
\put(6.1285,1.9){\emmv}
\put(6.1285,1.8625){\emln}
\put(6.4855,1.9){\emmv}
\put(6.4855,1.844){\emln}
\put(6.3605,1.619){\makebox(0,0)[lb]{\tmls 48}}
\put(6.843,1.9){\emmv}
\put(6.843,1.8625){\emln}
\put(7.2,1.9){\emmv}
\put(7.2,1.844){\emln}
\put(7.075,1.619){\makebox(0,0)[lb]{\tmls 56}}
\special{em:linewidth 0.4pt}
\linethickness{0.4pt}
\put(2.2,1.9){\emmv}
\put(2.2,5.15){\emln}
\put(2.2,1.9){\emmv}
\put(2.1435,1.9){\emln}
\put(1.944,1.8375){\makebox(0,0)[lb]{\tmls 0}}
\put(2.2,2.132){\emmv}
\put(2.1625,2.132){\emln}
\put(2.2,2.3645){\emmv}
\put(2.1435,2.3645){\emln}
\put(1.944,2.302){\makebox(0,0)[lb]{\tmls 2}}
\put(2.2,2.5965){\emmv}
\put(2.1625,2.5965){\emln}
\put(2.2,2.8285){\emmv}
\put(2.1435,2.8285){\emln}
\put(1.944,2.766){\makebox(0,0)[lb]{\tmls 4}}
\put(2.2,3.0605){\emmv}
\put(2.1625,3.0605){\emln}
\put(2.2,3.293){\emmv}
\put(2.1435,3.293){\emln}
\put(1.944,3.2305){\makebox(0,0)[lb]{\tmls 6}}
\put(2.2,3.525){\emmv}
\put(2.1625,3.525){\emln}
\put(2.2,3.757){\emmv}
\put(2.1435,3.757){\emln}
\put(1.944,3.6945){\makebox(0,0)[lb]{\tmls 8}}
\put(2.2,3.9895){\emmv}
\put(2.1625,3.9895){\emln}
\put(2.2,4.2215){\emmv}
\put(2.1435,4.2215){\emln}
\put(1.819,4.159){\makebox(0,0)[lb]{\tmls 10}}
\put(2.2,4.4535){\emmv}
\put(2.1625,4.4535){\emln}
\put(2.2,4.6855){\emmv}
\put(2.1435,4.6855){\emln}
\put(1.819,4.623){\makebox(0,0)[lb]{\tmls 12}}
\put(2.2,4.918){\emmv}
\put(2.1625,4.918){\emln}
\put(2.2,5.15){\emmv}
\put(2.1435,5.15){\emln}
\put(1.819,5.0875){\makebox(0,0)[lb]{\tmls 14}}
\special{em:linewidth 0.4pt}
\linethickness{0.4pt}
\put(2.2805,2.2325){\emmv}
\put(2.3695,2.383){\emln}
\put(2.459,2.497){\emln}
\put(2.548,2.592){\emln}
\put(2.6375,2.6755){\emln}
\put(2.727,2.751){\emln}
\put(2.816,2.8205){\emln}
\put(2.9055,2.885){\emln}
\put(2.9945,2.9455){\emln}
\put(3.084,3.0025){\emln}
\put(3.173,3.057){\emln}
\put(3.2625,3.109){\emln}
\put(3.352,3.1585){\emln}
\put(3.441,3.2065){\emln}
\put(3.5305,3.2525){\emln}
\put(3.6195,3.2975){\emln}
\put(3.709,3.3405){\emln}
\put(3.798,3.3825){\emln}
\put(3.8875,3.4235){\emln}
\put(3.977,3.463){\emln}
\put(4.066,3.502){\emln}
\put(4.1555,3.54){\emln}
\put(4.2445,3.577){\emln}
\put(4.334,3.613){\emln}
\put(4.423,3.6485){\emln}
\put(4.5125,3.6835){\emln}
\put(4.602,3.7175){\emln}
\put(4.691,3.751){\emln}
\put(4.7805,3.784){\emln}
\put(4.8695,3.816){\emln}
\put(4.959,3.848){\emln}
\put(5.048,3.879){\emln}
\put(5.1375,3.91){\emln}
\put(5.227,3.9405){\emln}
\put(5.316,3.97){\emln}
\put(5.4055,3.9995){\emln}
\put(5.4945,4.0285){\emln}
\put(5.584,4.0575){\emln}
\put(5.673,4.0855){\emln}
\put(5.7625,4.1135){\emln}
\put(5.852,4.141){\emln}
\put(5.941,4.1685){\emln}
\put(6.0305,4.195){\emln}
\put(6.1195,4.222){\emln}
\put(6.209,4.248){\emln}
\put(6.298,4.274){\emln}
\put(6.3875,4.3){\emln}
\put(6.477,4.3255){\emln}
\put(6.566,4.3505){\emln}
\special{em:linewidth 0.4pt}
\linethickness{0.4pt}
\put(2.2805,2.0395){\emmv}
\put(2.3785,2.1925){\emln}
\put(2.45,2.2875){\emln}
\put(2.557,2.4155){\emln}
\put(3.093,2.912){\emln}
\put(3.9855,3.4485){\emln}
\put(5.325,4.268){\emln}
\put(6.6645,4.941){\emln}
\special{em:linewidth 0.4pt}
\linethickness{0.4pt}
\put(2.2805,2.0395){\makebox(0,0)[cc]{$\ast$}}
\put(2.3785,2.1925){\makebox(0,0)[cc]{$\ast$}}
\put(2.45,2.2875){\makebox(0,0)[cc]{$\ast$}}
\put(2.557,2.4155){\makebox(0,0)[cc]{$\ast$}}
\put(3.093,2.912){\makebox(0,0)[cc]{$\ast$}}
\put(3.9855,3.4485){\makebox(0,0)[cc]{$\ast$}}
\put(5.325,4.268){\makebox(0,0)[cc]{$\ast$}}
\put(6.6645,4.941){\makebox(0,0)[cc]{$\ast$}}
\special{em:linewidth 0.4pt}
\linethickness{0.4pt}
\put(4.4765,4.6345){\makebox(0,0)[lb]{\tmls {\Huge Ar}}}
\special{em:linewidth 0.4pt}
\linethickness{0.4pt}
\put(6.785,2.0755){\makebox(0,0)[lb]{\tmls {\Large T}}}
\special{em:linewidth 0.4pt}
\linethickness{0.4pt}
\put(1.5975,4.7765){\Large\makebox(0,0)[cc]{$\lambda$}}
\special{em:linewidth 0.4pt}
\linethickness{0.4pt}
\put(2.2,5.15){\emmv}
\put(7.2,5.15){\emln}
\special{em:linewidth 0.4pt}
\linethickness{0.4pt}
\put(7.2,1.9){\emmv}
\put(7.2,5.15){\emln}
\end{picture}
\caption{Temperature dependence of thermal conductivity $\lambda$ of neutral
{\sf Ar} at $\Delta=0.0125$ ($n=6.074\cdot10^{20}$ cm$^{-3}$). The legend is 
the same as for figure \protect\ref{f1}, but experimental data were taken 
from \protect\cite{c25}. The transition relations to dimensional data read 
the same as for figure \ref{f2}.}
\label{f3}
\vspace*{5ex}
%\end{figure}
%\newpage
%
%
%
%\begin{figure}[h]
\unitlength=1in
\special{em:linewidth 0.4pt}
\linethickness{0.4pt}
\def\emmv{\special{em:moveto}}
\def\emln{\special{em:lineto}}
\def\tmls{\normalsize}
\begin{picture}(5.6025,3.531)(1.5975,1.619)
\special{em:linewidth 0.4pt}
\linethickness{0.4pt}
\put(2.2,1.9){\emmv}
\put(7.2,1.9){\emln}
\put(2.2,1.9){\emmv}
\put(2.2,1.844){\emln}
\put(2.1375,1.619){\makebox(0,0)[lb]{\tmls 0}}
\put(2.6165,1.9){\emmv}
\put(2.6165,1.8625){\emln}
\put(3.0335,1.9){\emmv}
\put(3.0335,1.844){\emln}
\put(2.971,1.619){\makebox(0,0)[lb]{\tmls 6}}
\put(3.45,1.9){\emmv}
\put(3.45,1.8625){\emln}
\put(3.8665,1.9){\emmv}
\put(3.8665,1.844){\emln}
\put(3.7415,1.619){\makebox(0,0)[lb]{\tmls 12}}
\put(4.2835,1.9){\emmv}
\put(4.2835,1.8625){\emln}
\put(4.7,1.9){\emmv}
\put(4.7,1.844){\emln}
\put(4.575,1.619){\makebox(0,0)[lb]{\tmls 18}}
\put(5.1165,1.9){\emmv}
\put(5.1165,1.8625){\emln}
\put(5.5335,1.9){\emmv}
\put(5.5335,1.844){\emln}
\put(5.4085,1.619){\makebox(0,0)[lb]{\tmls 24}}
\put(5.95,1.9){\emmv}
\put(5.95,1.8625){\emln}
\put(6.3665,1.9){\emmv}
\put(6.3665,1.844){\emln}
\put(6.2415,1.619){\makebox(0,0)[lb]{\tmls 30}}
\put(6.7835,1.9){\emmv}
\put(6.7835,1.8625){\emln}
\put(7.2,1.9){\emmv}
\put(7.2,1.844){\emln}
\put(7.075,1.619){\makebox(0,0)[lb]{\tmls 36}}
\special{em:linewidth 0.4pt}
\linethickness{0.4pt}
\put(2.2,1.9){\emmv}
\put(2.2,5.15){\emln}
\put(2.2,1.9){\emmv}
\put(2.1435,1.9){\emln}
\put(1.944,1.8375){\makebox(0,0)[lb]{\tmls 0}}
\put(2.2,2.171){\emmv}
\put(2.1625,2.171){\emln}
\put(2.2,2.4415){\emmv}
\put(2.1435,2.4415){\emln}
\put(1.944,2.379){\makebox(0,0)[lb]{\tmls 8}}
\put(2.2,2.7125){\emmv}
\put(2.1625,2.7125){\emln}
\put(2.2,2.9835){\emmv}
\put(2.1435,2.9835){\emln}
\put(1.819,2.921){\makebox(0,0)[lb]{\tmls 16}}
\put(2.2,3.254){\emmv}
\put(2.1625,3.254){\emln}
\put(2.2,3.525){\emmv}
\put(2.1435,3.525){\emln}
\put(1.819,3.4625){\makebox(0,0)[lb]{\tmls 24}}
\put(2.2,3.796){\emmv}
\put(2.1625,3.796){\emln}
\put(2.2,4.0665){\emmv}
\put(2.1435,4.0665){\emln}
\put(1.819,4.004){\makebox(0,0)[lb]{\tmls 32}}
\put(2.2,4.3375){\emmv}
\put(2.1625,4.3375){\emln}
\put(2.2,4.6085){\emmv}
\put(2.1435,4.6085){\emln}
\put(1.819,4.546){\makebox(0,0)[lb]{\tmls 40}}
\put(2.2,4.879){\emmv}
\put(2.1625,4.879){\emln}
\put(2.2,5.15){\emmv}
\put(2.1435,5.15){\emln}
\put(1.819,5.0875){\makebox(0,0)[lb]{\tmls 48}}
\special{em:linewidth 0.4pt}
\linethickness{0.4pt}
\put(2.478,2){\emmv}
\put(2.6165,2.0275){\emln}
\put(2.7555,2.0545){\emln}
\put(2.8945,2.0835){\emln}
\put(3.0335,2.1155){\emln}
\put(3.172,2.15){\emln}
\put(3.311,2.1885){\emln}
\put(3.45,2.2315){\emln}
\put(3.589,2.2785){\emln}
\put(3.728,2.331){\emln}
\put(3.8665,2.388){\emln}
\put(4.0055,2.4505){\emln}
\put(4.1445,2.519){\emln}
\put(4.2835,2.593){\emln}
\put(4.422,2.673){\emln}
\put(4.561,2.7595){\emln}
\put(4.7,2.852){\emln}
\put(4.839,2.951){\emln}
\put(4.978,3.057){\emln}
\put(5.1165,3.1695){\emln}
\put(5.2555,3.2895){\emln}
\put(5.3945,3.4165){\emln}
\put(5.5335,3.5505){\emln}
\put(5.672,3.692){\emln}
\put(5.811,3.841){\emln}
\put(5.95,3.9975){\emln}
\put(6.089,4.162){\emln}
\put(6.228,4.3345){\emln}
\special{em:linewidth 0.4pt}
\linethickness{0.4pt}
\put(2.478,1.9455){\emmv}
\put(2.6165,1.9605){\emln}
\put(2.7555,1.975){\emln}
\put(2.8945,1.9875){\emln}
\put(3.0335,1.999){\emln}
\put(3.172,2.038){\emln}
\put(3.311,2.097){\emln}
\put(3.45,2.178){\emln}
\put(3.589,2.282){\emln}
\put(3.728,2.432){\emln}
\put(3.8665,2.648){\emln}
\put(4.0055,2.829){\emln}
\put(4.1445,3.01){\emln}
\put(4.2835,3.1045){\emln}
\put(4.422,3.144){\emln}
\put(4.561,3.152){\emln}
\put(4.7,3.207){\emln}
\put(4.839,3.341){\emln}
\put(4.978,3.435){\emln}
\put(5.1165,3.561){\emln}
\put(5.2555,3.7345){\emln}
\put(5.3945,3.9155){\emln}
\put(5.5335,4.0885){\emln}
\put(5.672,4.2385){\emln}
\put(5.811,4.372){\emln}
\put(6.089,4.5925){\emln}
\put(6.3665,4.8605){\emln}
\special{em:linewidth 0.4pt}
\linethickness{0.4pt}
\put(2.478,1.9455){\makebox(0,0)[cc]{$\ast$}}
\put(2.8945,1.9875){\makebox(0,0)[cc]{$\ast$}}
\put(3.311,2.097){\makebox(0,0)[cc]{$\ast$}}
\put(3.728,2.432){\makebox(0,0)[cc]{$\ast$}}
\put(4.1445,3.01){\makebox(0,0)[cc]{$\ast$}}
\put(4.561,3.152){\makebox(0,0)[cc]{$\ast$}}
\put(4.978,3.435){\makebox(0,0)[cc]{$\ast$}}
\put(5.3945,3.9155){\makebox(0,0)[cc]{$\ast$}}
\put(5.811,4.372){\makebox(0,0)[cc]{$\ast$}}
\special{em:linewidth 0.4pt}
\linethickness{0.4pt}
\put(4.3225,4.492){\makebox(0,0)[lb]{\tmls {\Huge Ar$^+$}}}
\special{em:linewidth 0.4pt}
\linethickness{0.4pt}
\put(1.5975,4.7765){\Large\makebox(0,0)[cc]{$\lambda$}}
\special{em:linewidth 0.4pt}
\linethickness{0.4pt}
\put(6.785,2.0755){\makebox(0,0)[lb]{\tmls {\Large T}}}
\special{em:linewidth 0.4pt}
\linethickness{0.4pt}
\put(2.2,5.15){\emmv}
\put(7.2,5.15){\emln}
\special{em:linewidth 0.4pt}
\linethickness{0.4pt}
\put(7.2,1.9){\emmv}
\put(7.2,5.15){\emln}
\end{picture}
\caption{Temperature dependence of thermal conductivity $\lambda$ of 
once-ionized {\sf Ar} at $\Delta=0.0126$ ($n=6.123\cdot10^{20}$ cm$^{-3}$).
The legend is the same as for figure \protect\ref{f1}. The transition 
relations to dimensional data read: $\lambda_{dim}=\lambda\cdot10^{-1}\ 
Wt/(m\cdot K)$, $T_{dim}=T\cdot10^{3}\ K$.}
\label{f4}
\end{figure}
\end{document}